\documentclass[conference]{IEEEtran}
\IEEEoverridecommandlockouts
\usepackage{cite}
\usepackage{amsmath,amssymb,amsfonts}
\usepackage{algorithmic}
\usepackage{graphicx}
\usepackage{textcomp}
\usepackage{xcolor}
\usepackage{comment}
\usepackage{placeins}
\usepackage{float}

\usepackage{fancybox, subcaption, tikz, multirow, bm}
\newcommand{\fone}{$F_1$}
\newcommand{\foneb}{\bm{\fone}}
\newcommand{\rowgray}{gray!16}

\usepackage[table]{xcolor}
\usepackage{booktabs}
\usepackage[tt=false]{libertine}

\definecolor{myblue}{HTML}{0000EE}
\newcommand{\fakelink}[1]{%
  \textcolor{myblue}{{#1}}%
}

\def\BibTeX{{\rm B\kern-.05em{\sc i\kern-.025em b}\kern-.08em
    T\kern-.1667em\lower.7ex\hbox{E}\kern-.125emX}}
\begin{document}

\title{MultiPhishGuard: An Explainable and Adaptive Multi-Agent LLM System for Phishing Email Detection}

\author{

\IEEEauthorblockN{
Yinuo Xue, Eric Spero, Meng Wai Woo, Wei Gao, Giovanni Russello
}
\IEEEauthorblockA{
The University of Auckland, Auckland, New Zealand\\
yxue579@aucklanduni.ac.nz\\
\{eric.spero, wai.woo, w.gao, g.russello\}@auckland.ac.nz
}

}

\maketitle

\begin{abstract}
Phishing email detection faces significant challenges due to evolving adversarial tactics and heterogeneous attack patterns. Traditional approaches, such as rule-based filters and denylists, often struggle to keep pace, leading to missed detections and security risks. While machine learning methods have improved detection performance, they remain limited in adapting to novel and rapidly changing phishing strategies. 

We present MultiPhishGuard, an LLM-based multi-agent detection framework with learned coordination across specialized agents. The system consists of five cooperative agents (text, URL, metadata, explanation simplifier, and adversarial agents), with agent contributions dynamically weighted using Proximal Policy Optimization. To address emerging threats, the framework incorporates an adversarial training loop in which an LLM-based agent generates subtle, context-aware email variants to expose potential model weaknesses and improve robustness to ambiguous phishing cases.

Experimental evaluations on public datasets show that MultiPhishGuard achieves stronger performance than established baselines, including Chain-of-Thought prompting and single-agent variants, as supported by ablation studies and comparative analyses. The system achieves an accuracy of 97.89\%, with a false positive rate of 2.73\% and a false negative rate of 0.20\%. In addition, an explanation simplifier agent transforms technical model outputs into plain-language rationales intended for human users.

Overall, these results suggest that multi-agent LLM architectures with adaptive coordination and adversarial training represent a promising direction for phishing email detection.
\end{abstract}

\begin{IEEEkeywords}
Phishing Email Detection, Large Language Model, Multi-Agent System, Reinforcement Learning, Adversarial Training
\end{IEEEkeywords}

\section{Introduction}
Phishing remains one of the most persistent and damaging threats in cybersecurity, serving as a primary vector for data breaches and financial losses. The Anti-Phishing Working Group (APWG) observed 853,244 phishing attacks in the fourth quarter of 2025 \cite{APWG-Unifying}. Phishing attacks have grown increasingly sophisticated over recent years \cite{asiri2024phishingrtds}. These attacks have evolved from simple, deceptive messages to social engineering \cite{longtchi2024internet}, spear-phishing \cite{hazell2023spear}, and even AI-driven content generation \cite{roy2024chatbots,Proofprint-2025} to closely mimic legitimate communications. According to Verizon’s 2025 Data Breach Investigations Report \cite{Verizon-2025}, phishing remains a top breach cause, with many attacks bypassing traditional filters. Together, these findings underscore the need for adaptive, resilient phishing detection systems to handle diverse attack strategies.

Traditional phishing detection methods have relied on rule-based filters and denylists \cite{liu2005detecting}, which are limited by their inability to keep pace with evolving attacker strategies such as domain spoofing, dynamic URL obfuscation, and context-aware social engineering. Static machine learning models, while more adaptable, often depend on predefined features and historical data, making them ineffective against novel or subtle threats. Deep learning approaches, including CNNs, RNNs, and pre-trained models like BERT \cite{devlin2019bert}, have improved detection accuracy by capturing contextual language cues. However, they tend to focus primarily on email text or URLs, overlooking other modalities, and their ``black-box'' nature hinders interpretability. More recently, LLM-based systems have been proposed to analyze phishing emails using natural language understanding, showing promise in identifying nuanced patterns across various attack types \cite{koide2024chatspamdetector,rashid2025llms,asfour2023harnessing}. Despite their success, most LLM approaches remain single-agent architectures that produce binary outputs without transparent reasoning. They also lack adaptability \cite{roy2024chatbots}, as they are not optimized to learn from advanced attacks, such as spear phishing \cite{birthriya2025detection}. Crucially, phishing emails exhibit heterogeneous cues spanning linguistic manipulation in message content, infrastructural signals in URLs, and structural anomalies in headers and metadata, which are rarely modeled jointly by existing systems. These limitations underscore the need for a multi-modal, explainable, and adaptive detection framework that integrates diverse data sources and provides users with interpretable, evidence-based decisions.

\begin{figure}
    \centering
    \includegraphics[width=\linewidth]{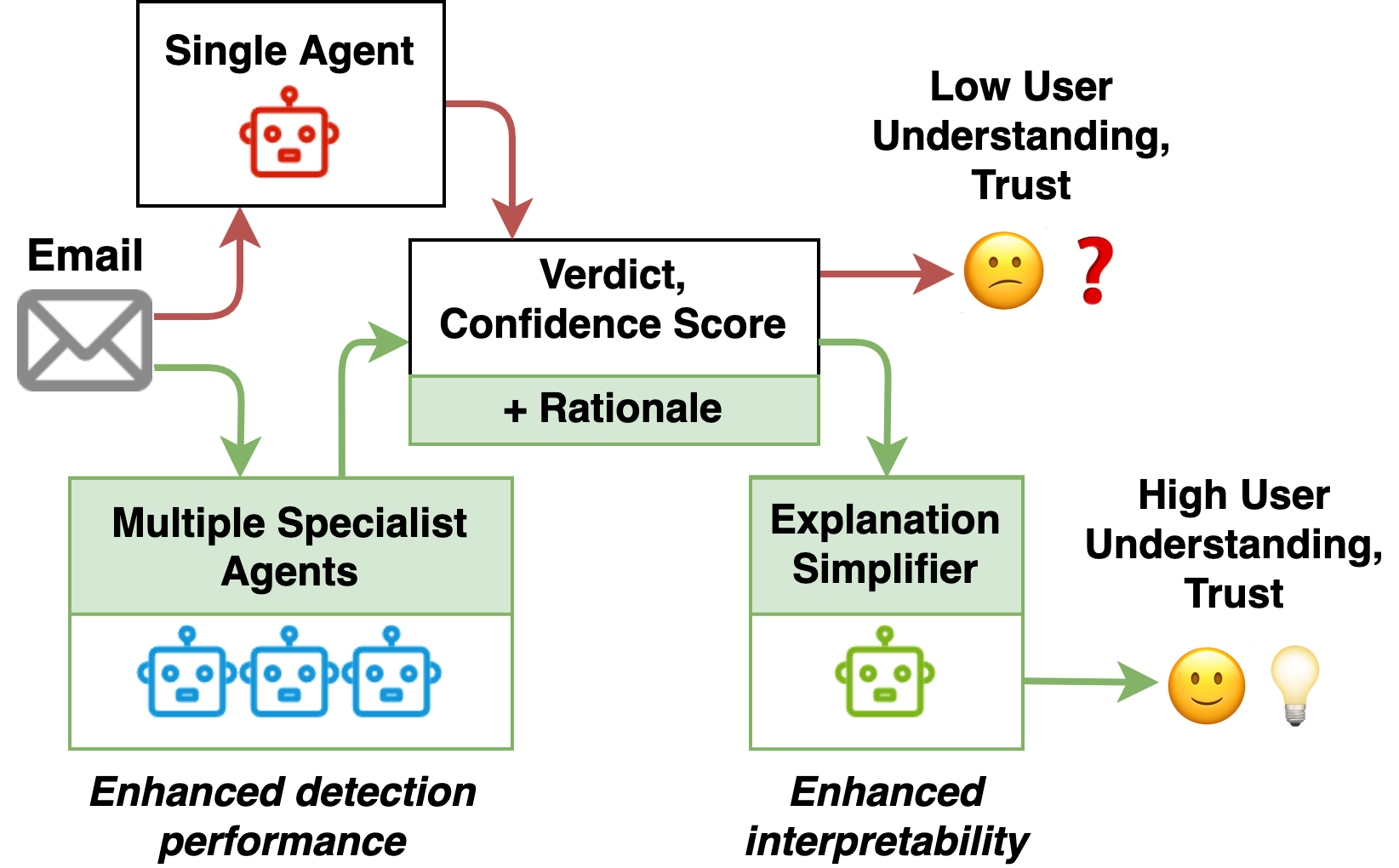}
\caption{Overview of the MultiPhishGuard detection pipeline. Incoming emails are processed by multiple specialized LLM agents that analyze different aspects of the message (metadata, text, and URLs). Each agent produces a prediction, confidence score, and rationale. An explanation simplifier agent then consolidates these outputs into a single plain-language explanation.}
    \label{fig:high-level}
\end{figure}

To address these limitations, we propose MultiPhishGuard, an LLM-based multi-agent system that explores the use of specialized, interacting agents for phishing email detection. The design is motivated by the observation that phishing emails exhibit heterogeneous cues across message content, URLs, and metadata, which naturally lend themselves to modular and specialized analysis. Unlike conventional ensemble methods, MultiPhishGuard employs interacting, specialized agents that generate structured rationales and adapt their coordination strategy during training. As illustrated in Figures \ref{fig:high-level} and \ref{fig: model_architecture}, our approach leverages LLMs not only to examine email content but also to scrutinize embedded URLs and metadata, enabling the system to reason over multiple aspects of an email in parallel. By distributing specialized tasks among distinct agents, each focused on a specific aspect of the email, the system dynamically fuses its outputs through a reinforcement learning mechanism that learns to weight agent contributions during integration. During training, adversarially generated variants of both phishing and legitimate emails are used to encourage robustness to subtle and ambiguous cases. Our evaluation indicates that this design shows promising performance in reducing false positives and handling challenging and ambiguous phishing cases under the datasets considered. Our main contributions are as follows:
\begin{itemize}
    \item We develop MultiPhishGuard, a novel LLM-based multi-agent system for phishing email detection that explicitly decomposes phishing analysis across specialized agents operating on email message body, embedded URLs, and metadata to capture malicious cues that traditional, single-modality detectors may miss. To the best of our knowledge, MultiPhishGuard is the first to utilize multiple LLM agents to deal with phishing emails. We also include our code as supplementary material for reproducibility.
    \item We incorporate adversarial training as a core component of the framework by deploying an adversarial agent that generates subtle variants of both phishing and legitimate emails. This approach is intended to improve robustness to subtle and ambiguous cases.
    \item We introduce an explanation simplifier agent that consolidates outputs from the text, URL, and metadata agents into concise, jargon-free rationales designed to support understanding by non-expert users. 
    \item We evaluated our model on six public datasets, achieving a 95.88\% \fone score, indicating robust performance across multiple public phishing datasets.
\end{itemize}

\section{Background and Related Work}

\subsection{Phishing Email Detection Methods.}
Traditional phishing detection methods initially relied on heuristic rules \cite{li2020detection}, denylists \cite{sharifi2008phishing}, and signature-based filtering \cite{liu2005detecting}. With the advancement of machine learning, ML algorithms have increasingly been applied to detect phishing emails. For example, \cite{cidon2019high} converts textual content into TF-IDF feature representations and trains classifiers such as random forests. With the emergence of deep learning, researchers have advanced these approaches by employing neural networks that automatically learn complex patterns from emails. For instance, researchers reformulated phishing email detection as text classification problems by leveraging pre-trained language models like BERT to identify phishing emails \cite{otieno2023application}. Despite these advances, key limitations remain in addressing evolving phishing strategies. The existing methods usually focus on a single modality and lack the adaptability to counter evolving phishing tactics rapidly \cite{kim2022phishing, haq2024detecting, verma2013semantic, ma2009beyond}. In contrast, our MultiPhishGuard fills these gaps by utilizing a multi-agent system to integrate diverse modalities to capture the phishing cues in emails.

\subsection{Heterogeneous Phishing Cues.}
Phishing emails contain multiple distinct types of cues that benefit from different forms of analysis. Phishing detection research commonly decomposes emails into headers and metadata, embedded URLs, and message body content, with each component yielding distinct structural, infrastructural, or linguistic cues and motivating specialized detection approaches \cite{almomani2013survey}. Recent work has shown that stacking ensemble approaches combining classifiers trained on heterogeneous email features (headers, URLs, and message content) can outperform single-model phishing detectors \cite{connolly2025effective}.

\subsection{Language Models for Phishing Detection.}
Phishing emails rely heavily on linguistic manipulation, contextual framing, and social engineering strategies that are difficult to capture using fixed feature representations alone. Large language models (LLMs) have been shown to effectively model semantic content, pragmatic intent, and contextual relationships in natural language tasks \cite{brown2020fewshot,wei2022chain}. These capabilities make LLMs well-suited for reasoning over subtle inconsistencies and deceptive cues that span message content, URLs, and metadata. As a result, LLMs provide a natural foundation for phishing detection systems that must integrate heterogeneous cues and adapt to evolving attack strategies.

\subsection{LLM-Based Multi-Agent Systems.}

Multi-agent systems have long been studied as a means of tackling complex tasks through decomposition into specialized subtasks. Foundational work by Shoham and Leyton-Brown \cite{shoham2008multiagent} formalizes how autonomous agents coordinate to achieve collective goals, while subsequent work in multi-agent reinforcement learning \cite{lowe2017multi} shows that such coordination improves performance in dynamic environments. This intuition aligns with ensemble learning, where combining multiple models improves predictive performance \cite{mienye2022ensemble}. Recent work extends this principle to large language models, where multi-agent systems leveraging role specialization have been shown to outperform single-agent approaches on complex tasks \cite{wu2024autogen,li2023camel,park2023generative,qiao2025thematic}.

These results suggest that aggregating specialized analyses is effective for complex problems. This is particularly relevant for phishing detection, where headers, URLs, and message content exhibit distinct structures and attack signals. Building on these insights, MultiPhishGuard integrates multiple LLM-based agents to enhance phishing email detection.

\subsection{Adaptive Decision-Making via Reinforcement Learning.}
Reinforcement learning (RL) enables systems to adaptively optimize decisions in dynamic environments by continuously adjusting parameters based on feedback. Foundational work by Sutton and Barto \cite{sutton1998reinforcement} established the theoretical basis for RL, which has since driven advancements across domains. Early models like Deep Q-Networks (DQN) \cite{mnih2015human} and Deep Deterministic Policy Gradient (DDPG) \cite{lillicrap2015continuous} showed how agents can learn effective policies through interaction, while later methods such as Proximal Policy Optimization \cite{schulman2017proximal} improved stability and efficiency. These approaches offer a key advantage over static models by dynamically tuning parameters in real-time. In MultiPhishGuard, RL is integrated into a multi-agent framework to dynamically adjust agent weights based on the characteristics of each email, enhancing coordinated decision-making and improving adaptability in the face of evolving phishing threats.

\subsection{Adversarial Training.}
Adversarial training has emerged as a key strategy for enhancing the robustness of machine learning models, especially in cybersecurity applications. The seminal work by Goodfellow et al. \cite{goodfellow2014explaining} demonstrated that incorporating adversarial examples—carefully crafted inputs designed to mislead a model—into training can improve resistance to such attacks. These examples can also simulate evolving attacker tactics to expose vulnerabilities in static models \cite{huang2011adversarial, biggio2018wild}.

In cybersecurity, adversarial training helps improve robustness and reduce false negatives by preparing systems to recognize sophisticated evasion techniques. By exposing the model to adversarially generated variants that reflect realistic phishing strategies, the approach encourages the system to adapt to subtle and evolving attack patterns rather than relying on static cues. Our model utilizes an adversarial agent to generate both phishing and legitimate email variants based on real-world emails, enhancing robustness to evolving threats.

\subsection{Interpretability.}
Interpretability---the extent to which a human can understand the basis of a model’s decision---is increasingly important for deploying machine learning systems in real-world settings. Model explanations can help users understand system behavior, identify failure modes, and support more informed decision-making \cite{ribeiro2016should}. Regulatory frameworks (e.g., GDPR’s “right to explanation”) also place growing emphasis on transparency, making interpretability a practical consideration in many deployment contexts \cite{goodman2017european, EU_AI_Act_2024}.


\section{MultiPhishGuard}
\begin{figure*}
    \centering
    \includegraphics[width=\linewidth]{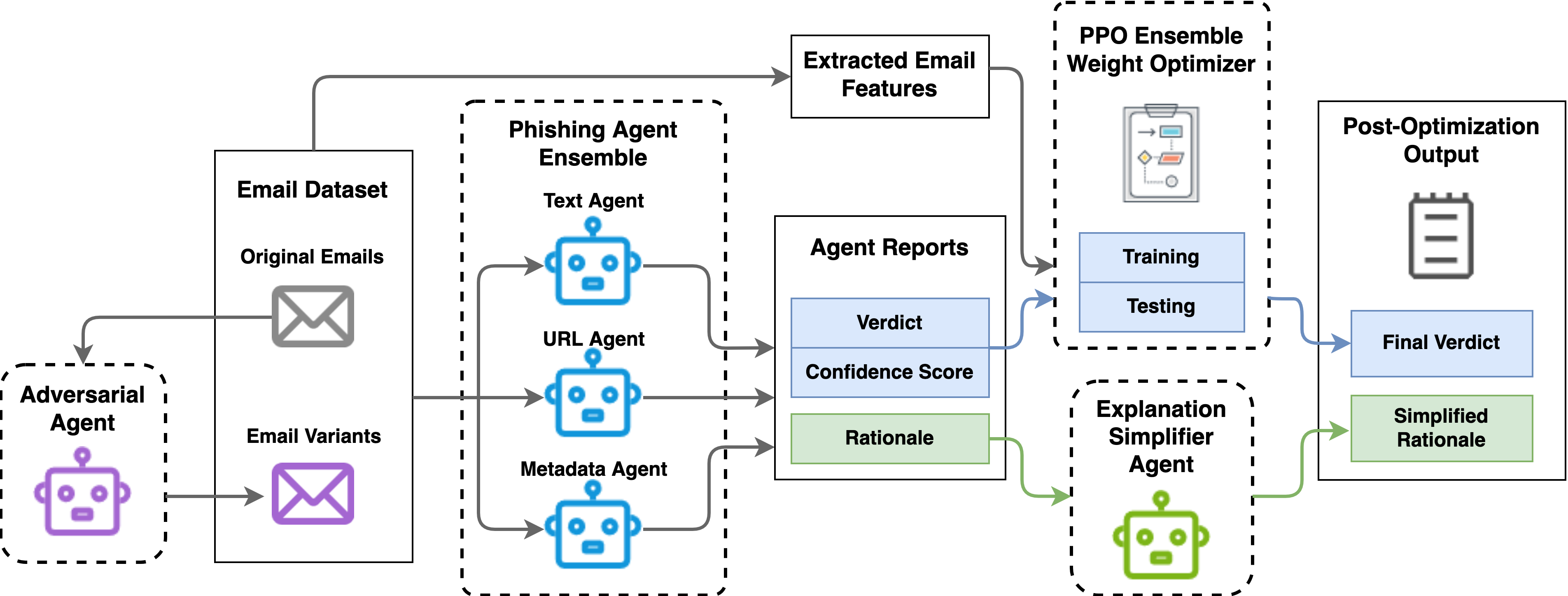}
    \caption{\textbf{MultiPhishGuard Architecture}. The Adversarial Agent generates subtle variants of both phishing and legitimate emails. For each email, three specialized agents---the Text Agent (analyzes the message body), URL Agent (inspects links), and Metadata Agent (evaluates headers and sender authentication records)---produce a verdict, confidence score, and rationale. These outputs, along with extracted email features, are fed into a Proximal Policy Optimization (PPO) module. During training, PPO iteratively updates agent weights to improve detection accuracy; the learned weights are applied during testing. Finally, a Rationale Simplifier Agent consolidates the rationales into concise, user-friendly explanations.
    }
    \label{fig: model_architecture}
\end{figure*}

Our proposed methodology leverages an LLM-based multi-agent system to enhance phishing email detection through a dynamic, adaptive, and explainable approach. As shown in Figure \ref{fig: model_architecture}, the framework consists of several specialized agents, each focusing on distinct aspects of email analysis. By integrating multiple agents, dynamically adjusting their influence using reinforcement learning, and incorporating adversarial training, our model is designed to improve detection accuracy and robustness while providing clear, user-friendly explanations. This approach aims to address the limitations of static, single-modality detectors, offering a more robust defense against phishing threats.

\subsection{Basic Agents}

Our phishing detection framework is built upon an LLM-based multi-agent system, where each agent specializes in analyzing different parts of an email. Motivated by prior work on ensemble learning and LLM-based multi-agent systems, which shows that combining specialized components can improve performance on complex tasks, our approach decomposes phishing detection into multiple complementary analyses. Unlike single-agent approaches that target only one modality (e.g., text, URL, or metadata) \cite{kim2022phishing, haq2024detecting, verma2013semantic}, our system leverages multiple perspectives to support detection accuracy and robustness. Each agent operates independently, providing its own phishing verdict, confidence score, and reasoning. These outputs are then dynamically fused using reinforcement learning to produce a final classification. The agents are built using AutoGen \cite{wu2023autogen}. Our system comprises three main components: a text analysis agent, a URL analysis agent, and a metadata analysis agent.

\noindent\textbf{Text Analysis Agent:} The text agent leverages an LLM to thoroughly analyze the email body, identifying suspicious patterns, phishing keywords, and any textual indicators of malicious intent. The prompt used by the agent is illustrated in Figure \ref{fig: text agent prompt}.

\begin{figure}
    \centering
\begin{tikzpicture}[
  box/.style={
    minimum width=1cm,
    inner sep=6pt,
    align=left, 
    rounded corners=8pt,
    fill=black!5
  }
  ]
    \node[box] (A) {%
\begin{minipage}{\dimexpr\columnwidth-12pt}

\setlength{\parskip}{1.5ex}

\ttfamily\small

You are a cybersecurity expert specializing in phishing, with a particular focus on email text content. Your task is to examine the email body exclusively for phishing cues—such as abnormal language patterns, suspicious vocabulary, or any textual indicators of malicious intent. Do not analyze URLs or metadata, only focus on the email text. Provide your judgment on whether the email is `Phishing' or `Legitimate', along with a confidence score between 0 and 1 and a clear, concise explanation of your reasoning. Output your result in JSON format as: {`verdict': `Phishing' or `Legitimate', `confidence': 0-1, `reasons': `...'}

\end{minipage}
};
\end{tikzpicture}%
    \caption{Text Agent Prompt}
    \label{fig: text agent prompt}
\end{figure}

\noindent\textbf{URL Analysis Agent:} The URL agent extracts and scrutinizes all links in the email, checking for obfuscation, evaluating domain reputations, and detecting potential redirections to malicious sites. The prompt used by this agent is presented in Appendix \ref{prompts} (Figure \ref{fig: URL agent prompt}). The main difference between the prompts used by the URL and text agents is that the former is tailored to focus exclusively on suspicious URLs, while the latter analyzes only the textual content.

\noindent\textbf{Metadata Analysis Agent:} The metadata agent examines email headers, sender authentication records (e.g., SPF, DKIM, DMARC), and reply-to fields to identify anomalies that may signal phishing attempts. The corresponding prompt is shown in Appendix \ref{prompts} (Figure \ref{fig: metadata agent prompt}). Unlike the text and URL agents, which focus respectively on the email body and embedded links, the metadata agent is specifically instructed to analyze only the metadata.


Each agent processes the email independently and generates a structured output containing (1) a phishing vs. legitimate label, (2) a phishing possibility score, and (3) a rationale for its decision.

As shown above, we adopted the same prompt format for the Text Agent, URL Agent, and Metadata Agent. This design was chosen to precisely define each agent's scope, structure its reasoning, and generate machine-readable output---drawing on best practices from the prompt engineering literature.

The prompts begin with ``You are a cybersecurity expert specializing in phishing,'' reflecting the ``system message'' approach used in InstructGPT \cite{ouyang2022training}, which has been shown to significantly improve task adherence by clearly defining the model’s role and focus area. By explicitly restricting the analysis to the email body (``Only focus on the email text; do not analyze URLs or metadata''), we prevent cross-modal interference and promote agent specialization \cite{park2023generative}---an approach also supported in multi-agent frameworks for complex tasks. Additionally, requiring the output in JSON format (\texttt{\{`verdict': `Phishing', `confidence': 0.87, `rationale': `...'\}}) aligns with structured prompting principles recommended by Toolformer \cite{schick2023toolformer}, enabling consistent integration, streamlined automated evaluation, and reliable downstream parsing.

\subsection{Dynamic Weight Adjustment}
Our framework employs a reinforcement learning (RL)-based mechanism to adaptively weight the outputs of specialized agents based on the characteristics of each email. Instead of relying on fixed weights, the system learns a weighting policy that assigns agent contributions on a per-email basis. For each email, the text, URL, and metadata agents generate independent predictions along with confidence scores. The RL module additionally receives a vector of email-specific features extracted during preprocessing, including the number of URLs present, key phishing keywords in the text, sender domain reputation scores, authentication results (SPF/DKIM/DMARC), and the individual confidence scores provided by each agent.

Concretely, let each agent \( i \) produce a prediction probability \( p_i \) along with a confidence score, and let the weight assigned to that agent be \( w_i \) (with \( \sum_i w_i = 1 \)). The final phishing score \( y \) is computed as a weighted sum of the agents' predictions:

\[
y = \sum_{i=1}^{N} w_i \cdot p_i.
\]

To optimize these weights dynamically, our system employs Proximal Policy Optimization (PPO), an efficient policy gradient method known for its stability and reliability in updating policies. We chose PPO over other reinforcement learning algorithms for several key reasons. First, PPO’s clipped surrogate objective provides a simple yet effective way to enforce a ``trust region'' on policy updates, ensuring that each gradient step remains within a safe bound and preventing the large, destabilizing policy swings that can occur with vanilla policy gradients or Advantage Actor Critic (A2C). Second, PPO naturally handles continuous action spaces---such as our weight vectors---without requiring complex discretization or Q-function approximations, unlike DQN or DDPG. Third, compared to more complex trust-region methods like TRPO, PPO is far easier to implement and tune, with fewer hyperparameters and lower computational overhead. Finally, PPO has demonstrated strong empirical performance and sample efficiency across a wide range of continuous control and decision-making tasks, making it a robust, reliable choice for dynamically optimizing our multi-agent weighting scheme.

The PPO's clipped surrogate objective function is given by:
\[
L(\theta) = \mathbb{E}\left[ \min\left( r(\theta) \hat{A}, \text{clip}(r(\theta), 1 - \epsilon, 1 + \epsilon) \hat{A} \right) \right]
\]
where \( r(\theta) = \frac{\pi_\theta(w|x)}{\pi_{\theta_{\text{old}}}(w|x)} \) is the probability ratio between the current and previous policies, \( \hat{A} \) is the estimated advantage, and \( \epsilon \) is a small hyperparameter to limit policy updates. The ``clip'' is a mechanism that limits how much the probability ratio \(r(\theta)\) can change during an update, ensuring that each policy update remains within a specified range. Concretely, the clip function takes the ratio \(r(\theta)\) and forces it to lie between \(1 - \epsilon\) and \(1 + \epsilon\). This prevents large policy changes by effectively ``clipping'' the advantage, \(\hat{A}\) when the new policy deviates excessively from the old policy. This formulation ensures that the weight updates remain within a bounded range, promoting stable learning.

In our implementation, the RL module treats the weights \( w = [w_1, w_2, \ldots, w_N] \) as actions drawn from a policy \( \pi_\theta(w|x) \), where \( x \) represents email-specific features (e.g., number of URLs and key phishing keywords in the text). The objective is to maximize the expected reward \( E[r] \), and the reward function is the accuracy of the final classification. Through iterative feedback and reward optimization, PPO allows the system to refine its weighting strategy, continuously adapting to subtle and evolving phishing patterns.

This dynamic weighting strategy is designed to enhance detection accuracy and reduce false positives by tailoring the influence of each modality based on context. For example, when an email contains suspicious links, the model may prioritize URL analysis, whereas it may focus on metadata if the sender information appears inconsistent. The PPO-based RL module continuously optimizes these weight adjustments, helping the detection framework remain robust across a range of phishing tactics. Consequently, our approach provides an adaptive detection framework that can refine its behavior as phishing strategies evolve.

\subsection{Explanation Simplifier Agent}

To support user understanding and trust in phishing email detection, our framework incorporates an explanation simplifier agent. Unlike traditional models that output only binary labels or confidence scores with little explanation, our system generates clear, user-friendly rationales summarizing why an email was flagged as phishing. This feature is useful for non-expert users, security analysts, and organizations that require interpretable results for decision-making and cybersecurity awareness training.

Given that our multi-agent system analyzes email text, URLs, and metadata separately, each agent produces an independent decision with its own confidence score and technical reasoning. Presenting these raw rationales directly can be overwhelming for non-expert users. To address this, the explanation simplifier agent---illustrated in Figure \ref{fig: explanation simplifier agent prompt}---aggregates the outputs, extracts key insights, and synthesizes them into a single, coherent explanation.

\begin{figure}
\centering
\begin{tikzpicture}[
  box/.style={
    minimum width=1cm,
    inner sep=6pt,
    align=left, 
    rounded corners=8pt,
    fill=black!5
  }
  ]
    \node[box] (A) {%
\begin{minipage}{\dimexpr\columnwidth-12pt}

\setlength{\parskip}{1.5ex}

\ttfamily\small
You are an expert in cybersecurity with deep expertise in phishing. Your task is to take the detailed technical explanations provided by the three specialized agents (text, URL, and metadata) for why an email is classified as phishing or legitimate, and synthesize them into one coherent, reliable, and complete explanation written in plain, everyday language. Ensure that your explanation is truthful, meaningful, and based solely on factual evidence—do not include any fabricated details. Avoid technical jargon, simplify complex concepts, and provide clear, concise reasons for the classification that accurately reflect the underlying data.
\end{minipage}
      };
\end{tikzpicture}%
\caption{Explanation Simplifier Agent Prompt}
\label{fig: explanation simplifier agent prompt}
\end{figure}

The explanation simplifier is prompted to produce concise, plain-language summaries grounded in the agents’ outputs, while avoiding unnecessary technical detail.

The final explanation produced by the agent: 
\textbf{(1)} synthesizes multi-agent insights into a single, coherent response; 
\textbf{(2)} reduces technical jargon to improve accessibility for non-expert users; 
\textbf{(3)} highlights key phishing indicators, such as suspicious sender details, deceptive language, or malicious links.

To accommodate users with different technical backgrounds, we also include an Expert Mode that provides more detailed explanations, including indicators of compromise and email header analysis. Evaluating the usability of these explanations across user groups is left for future work.

\subsection{Adversarial Training Module}
To improve robustness, we introduce an adversarial training module that exposes the model to challenging and evasive examples. This module leverages an adversarial agent—a GPT-4o-based large language model—to generate nuanced variants of both phishing and legitimate emails. These variants are designed to mimic evasion strategies while remaining within ethical and controlled boundaries. Training on such examples is intended to improve robustness to real-world threats and increase sensitivity to diverse phishing patterns.

Adversarial training is implemented as an iterative process in which the adversarial agent acts as a generator, crafting email variants that aim to evade detection. Unlike traditional approaches that rely on small perturbations, our agent modifies emails to reflect realistic phishing tactics. These adversarial examples are evaluated by the multi-agent detection model, which acts as a discriminator. Misclassifications are used as feedback to refine the agent’s strategy, forming a feedback loop that is intended to improve robustness to challenging and evolving phishing strategies.

The adversarial agent generates phishing emails that resemble authentic communications while avoiding common signatures. It modifies existing emails, creates new deceptive variants, and refines them based on model feedback. In parallel, it generates legitimate emails that mimic phishing traits—without becoming malicious—to test the model’s ability to make fine-grained distinctions. This dual-generation process is intended to improve performance in ambiguous cases.

Formally, let \( f(x; \theta) \) denote the detection model, which outputs the probability \( p \) of an email \( x \) being phishing. The adversarial agent generates inputs \( x_{\text{adv}} \) designed to challenge the model. Unlike traditional adversarial training, where \( x_{\text{adv}} = x + \delta \) with \( \|\delta\| \leq \epsilon \), our approach synthesizes context-aware adversarial examples that simulate real-world evasion techniques.

As shown in Figure \ref{fig: adversarial agent prompt}, the agent employs five transformation strategies: Synonym Substitution \cite{samanta2017towards}, Sentence Rewriting \cite{xu2021r}, Content Modification, Homoglyph Replacement \cite{almuhaideb2022homoglyph}, and Polymorphic Variation \cite{chauhan2021polymorphic}. These transformations preserve semantic intent while introducing variations designed to challenge detection. An example is provided in Appendix \ref{adversarial_email}.

To mitigate misuse risks, adversarial emails generated during our experiments are not publicly released and are used solely for internal testing and model refinement.

Overall, this adversarial training framework is designed to improve robustness to common evasion strategies and support adaptation to evolving phishing tactics.

\section{Experiments}
We employ GPT-4o\footnote{https://platform.openai.com/docs/models/gpt-4o} as our LLM agent, with JSON mode activated to ensure a consistent and easy-to-evaluate output format. We have selected GPT-4o for its stable API access and consistently reliable performance. Its architecture integrates seamlessly with AutoGen’s multi-agent framework, facilitating the development of collaborative AI agents. A series of experiments was conducted to evaluate MultiPhishGuard's effectiveness in both identifying phishing emails and generating well-reasoned explanations. 


\subsection{Datasets} \label{eval:dataset}
We evaluate our proposed LLM-based multi-agent system using six widely recognized datasets: the Nazario phishing corpus \cite{Nazario2005The}, the Enron-Spam dataset \cite{metsis2006spam}, the TREC 2007 public corpus \cite{Gordon2007TREC}, the CEAS 2008 public corpus \cite{2008CEAS}, the Nigerian Fraud dataset \cite{Radev2008CLAIR}, and the SpamAssassin public corpus \cite{Apache2006Spam}. These datasets encompass a diverse collection of phishing and legitimate emails, providing a diverse and comprehensive testbed for our detection framework.

The Nazario phishing corpus comprises Jose's self-reported phishing emails collected from 2005 to 2024. To ensure our model is tuned to recognize the most recent phishing indicators, we selected only the phishing emails from 2024, totaling 402 emails. Additionally, the Nigerian Fraud dataset contains over 2,500 emails dating from 1998 to 2007; we chose the most recent subset from 2007, which includes 577 emails. In contrast, the Enron-Spam dataset comprises both spam and ham emails. Since our study specifically targets phishing detection rather than general spam filtering, we used the ham emails as examples of legitimate messages, randomly sampling 1,500 emails for our experiments. To further demonstrate the model's transferability to other legitimate datasets, we randomly selected 500 ham emails each from the TREC 2007 and CEAS 2008 public corpora. In addition, we selected 500 ``hard ham'' emails from the SpamAssassin public corpus to ensure our model is capable of accurately classifying even the most challenging legitimate emails.

Overall, MultiPhishGuard was evaluated on 979 phishing emails and 3,000 legitimate emails---approximately 4,000 emails in total---with a phishing-to-legitimate ratio of roughly 1:3.

\subsection{Evaluation Metrics}\label{eval:metrics}

\subsubsection{Evaluation of Phishing Detection} 
To comprehensively evaluate the performance of our phishing detection system, we utilize several standard classification metrics, including Recall, Precision, Accuracy, \fone score, True Negative Rate (TNR), False Positive Rate (FPR), and False Negative Rate (FNR). These metrics provide insights into the effectiveness of the model in correctly identifying phishing emails while minimizing false detections. Explanations of these metrics can be found in Appendix \ref{classification_metrics}. By incorporating these evaluation metrics, we ensure a balanced assessment of the model's performance, focusing not only on detection accuracy but also on minimizing the risks of false positives (which disrupt users) and false negatives (which allow phishing attacks to succeed), which are critical for maintaining both security and user experience.

\subsubsection{Evaluation of Generated Rationale}
In addition to evaluating discriminative performance, we assess the quality of the explanations generated by the explanation simplifier agent using both automated metrics and human evaluation. Since interpretability plays a crucial role in user trust and decision-making, we measure the clarity, coherence, and readability of the generated explanations through the Perplexity \cite{jelinek1977perplexity}, Topic Coherence \cite{rosner2014evaluating}, and Flesch Reading Ease Score \cite{flesch1979write}. Further explanations about these metrics can be found in Appendix \ref{rationales_metrics}. By combining these metrics, we aim to ensure that our explanations are not only technically sound but also clear, user-friendly, and informative, ultimately improving trust and usability in phishing detection.

\subsection{Comparative Evaluation}

We perform a comparative evaluation of our proposed MultiPhishGuard against three alternative approaches---Chain-of-Thought, a single-agent model, and the RoBERTa-base baseline---on the datasets and evaluation metrics described in Sections \ref{eval:dataset} and \ref{eval:metrics}. 

Our primary analysis evaluates overall performance on the pooled dataset, as shown in Table \ref{tab:evaluation_detection}, with additional per-dataset results reported in Table \ref{tab:evaluation_detection_dataset} and Section \ref{sec:mcnemar_comp}.

\subsubsection{MultiPhishGuard}

The experiments were conducted within a framework that integrates multiple specialized agents (text, URL, and metadata) whose outputs are dynamically fused using a PPO-based reinforcement learning module. Our framework also incorporates adversarial training to continuously challenge the system and an explanation simplifier agent to provide clear, user-friendly rationales for its phishing/legitimate classifications.

In our experiments, MultiPhishGuard demonstrated strong performance on phishing detection tasks. Evaluated on a comprehensive dataset, the system achieved a recall of 99.80\% and a precision of 92.26\%, resulting in an overall accuracy of 97.89\% and an \fone score of 95.88\%. The true negative rate was 97.27\%, with a false positive rate of 2.73\% and a false negative rate of 0.20\%.

\begin{table*}
    \rowcolors{1}{white}{\rowgray}
    \centering
  \caption{Overall phishing detection performance across all datasets (all values in \%).}
    \label{tab:evaluation_detection}
    \begin{tabular}{lccccccc}
    \toprule
        \textbf{Approach} & \textbf{Recall} & \textbf{Precision} & \textbf{Accuracy} & \textbf{\foneb score} & \textbf{TNR} & \textbf{FPR} & \textbf{FNR} \\ 
        \midrule
        \textit{MultiPhishGuard} & \textbf{99.80} & \textbf{92.26} & \textbf{97.89} & \textbf{95.88} & \textbf{97.27} & \textbf{2.73} & \textbf{0.20} \\ 
        CoT & 99.08 & 64.93 & 86.60 & 78.45 & 82.53 & 17.47 & 0.92 \\ 
        Single Agent & 99.39 & 61.31 & 84.42 & 75.84 & 79.53 & 20.47 & 0.61 \\ 
        RoBERTa-base & 98.37 & 86.21 & 95.73 & 91.89 & 94.87 & 5.13 & 1.63 \\ 
        \bottomrule
    \end{tabular}
\end{table*}

\subsubsection{Chain-of-Thought} \label{CoT}

We compared a Chain-of-Thought (CoT) prompting approach against our proposed MultiPhishGuard system. The CoT method leverages the ability of LLMs to articulate intermediate reasoning steps before arriving at a final decision---a technique that has been shown to improve performance on complex reasoning tasks \cite{wei2022chain}. When applied to phishing detection, the CoT approach encourages the model to explain its thought process, potentially revealing subtle cues in the email content. While CoT prompting is effective in eliciting intermediate reasoning steps in certain tasks, it presents several challenges for phishing detection.

As shown in Table \ref{tab:evaluation_detection}, the CoT approach demonstrates a high recall rate of 99.08\%, indicating that it effectively identifies the majority of phishing emails. However, its precision is considerably lower at 64.93\%, largely due to a false positive rate of 17.47\% and a true negative rate of 82.53\%. Consequently, while the CoT method detects most phishing instances (with only a 0.92\% false negative rate), it also misclassifies a significant number of legitimate emails as phishing. This imbalance results in an overall accuracy of 86.60\% and an \fone score of 78.45\%, highlighting the method's tendency to generate noisy outputs. These results suggest that, although CoT is effective at flagging potential phishing emails, its high false positive rate may limit its practical reliability in real-world applications. These results may stem from the CoT methods generating inconsistent or excessively detailed reasoning paths, which can compromise the detection of nuanced, multi-modal phishing cues. 

The CoT approach exhibits several critical shortcomings in its explanations. As shown in Table \ref{tab:evaluation_reasons}, a perplexity score of 47 suggests that the generated text is relatively less fluent and predictable, which can hinder natural language understanding. Additionally, a topic coherence score of 0.28 indicates that the explanations lack semantic consistency, making it difficult for users to discern clear and logical reasoning. Furthermore, with a Flesch Reading Ease Score of only 21, the explanations are notably hard to read, implying that they are filled with technical jargon and complex sentence structures that impede accessibility. Collectively, these issues highlight that the CoT method struggles to provide explanations that are both coherent and user-friendly.

\subsubsection{Single-Agent Model} \label{SingleAgent}

We also compared a single-agent model with our proposed MultiPhishGuard system to evaluate their effectiveness in phishing detection. The single-agent approach relies on one LLM (GPT-4o) to process the entire email and generate a phishing verdict directly. While this method has been applied in prior studies and can yield satisfactory results in simpler scenarios, it typically focuses on a single modality. This limitation often leads to higher false positive rates and reduced robustness when facing sophisticated phishing tactics.

The single-agent approach achieves an impressive recall rate of 99.39\% and a false negative rate of 0.61\%, indicating that it is highly effective at detecting phishing emails. However, its precision is only 61.31\%, accompanied by a relatively high false positive rate of 20.47\% and a low true negative rate of 79.53\%. This means that while the system successfully identifies most phishing instances, it also incorrectly classifies a significant number of legitimate emails as phishing, leading to an overall accuracy of 84.42\% and an \fone score of 75.84\%. These issues highlight the inherent limitations of a single-agent model, which, by relying on a solitary modality, may fail to capture the diverse and nuanced cues present in complex, multi-modal phishing attacks.

\subsubsection{Baseline: RoBERTa-base} \label{Baseline}

Since RoBERTa-base \cite{liu2019roberta} outperformed other pre-trained language models \cite{roy2024chatbots}---including BERT-base \cite{devlin2019bert}, DistilBERT \cite{sanh2019distilbert}, ELECTRA \cite{clark2020electra}, DeBERTa \cite{he2020deberta}, and XLNet \cite{yang2019xlnet}---we chose it as the baseline for our phishing detection experiments. Widely used for natural language understanding tasks, RoBERTa-base is pre-trained on extensive corpora and optimized through dynamic masking strategies, making it particularly effective for text classification. To ensure a fair comparison, we used the same training settings as in \cite{roy2024chatbots}.

In our experiments, RoBERTa-base achieved a recall of 98.37\%, accuracy of 95.73\%, true negative rate of 94.87\%, and an \fone score of 91.89\%. While these results indicate strong overall detection performance, several issues remain. First, the precision of 86.21\% coupled with a false positive rate of 5.13\% suggests that a notable number of legitimate emails are incorrectly flagged as phishing, which could lead to unnecessary disruptions and reduced user trust. Also, the false negative rate of 1.63\% indicates that some phishing emails are still missed, posing potential security risks.

Another significant limitation of RoBERTa-base is its black-box nature; it does not provide interpretable reasoning or explanations behind its classifications. This lack of transparency makes it difficult for users and security analysts to understand why a particular email was labeled as phishing, hindering efforts to fine-tune detection criteria and improve overall system trustworthiness. These shortcomings highlight the need for a more comprehensive approach, such as our proposed MultiPhishGuard, which not only enhances detection accuracy but also delivers clear, user-friendly explanations for its decisions.

\subsubsection{Performance Across Datasets}
Table \ref{tab:evaluation_detection_dataset} reports performance broken down by dataset. While our primary analysis focuses on pooled results, the per-dataset breakdown shows a consistent pattern: MultiPhishGuard maintains near-perfect recall on phishing datasets while substantially reducing false positive rates across legitimate datasets compared to CoT and single-agent approaches. It also compares favorably to RoBERTa-base, achieving lower false positive rates across most datasets while maintaining comparable recall.

\begin{table*}
    \centering
    \rowcolors{5}{white}{\rowgray}
    \caption{Per-dataset classification performance across six email corpora. For phishing datasets, we report Recall (TPR) and FNR; for legitimate datasets, we report TNR and FPR (all values in \%).}
    \label{tab:evaluation_detection_dataset}
    \begin{tabular}{lrrrrrrrrrrrr}
    \toprule
    &\multicolumn{4}{c}{\textbf{Phishing}} & \multicolumn{8}{c}{\textbf{Legitimate}}\\\cmidrule(lr){2-5}\cmidrule(lr){6-13}
    & \multicolumn{2}{c}{\textbf{Nigerian}} & \multicolumn{2}{c}{\textbf{Nazario}} & \multicolumn{2}{c}{\textbf{Enron}} & \multicolumn{2}{c}{\textbf{SpamAssassin}} & \multicolumn{2}{c}{\textbf{CEAS-08}} & \multicolumn{2}{c}{\textbf{TREC-07}} \\\cmidrule(lr){2-3}\cmidrule(lr){4-5}\cmidrule(lr){6-7}\cmidrule(lr){8-9}\cmidrule(lr){10-11}\cmidrule(lr){12-13}
    \textbf{Approach} & \textbf{TPR} & \textbf{FNR} & \textbf{TPR} & \textbf{FNR} & \textbf{TNR} & \textbf{FPR} & \textbf{TNR} & \textbf{FPR} & \textbf{TNR} & \textbf{FPR} & \textbf{TNR} & \textbf{FPR}\\\midrule
    \textit{MultiPhishGuard} & \textbf{100.00} & \textbf{0.00} & \textbf{99.50} & \textbf{0.50} & \textbf{97.47} & \textbf{2.53} & \textbf{95.00} & \textbf{5.00} & \textbf{98.20} & \textbf{1.80} & \textbf{98.00} & \textbf{2.00}\\
    CoT & 99.83 & 0.17 & 98.01 & 1.99 & 80.20 & 19.80 & 73.40 & 26.60 & 88.80 & 11.20 & 92.40 & 7.60 \\
    Single Agent & \textbf{100.00} & \textbf{0.00} & 98.51 & 1.49 & 77.33 & 22.67 & 63.20 & 36.80 & 88.60 & 11.40 & 93.40 & 6.60 \\
    RoBERTa-base & 99.31 & 0.69 & 97.01 & 2.99 & 95.80 & 4.20 & 89.40 & 10.60 & 96.40 & 3.60 & 96.00 & 4.00 \\
    \bottomrule
    \end{tabular}
\end{table*}

\subsubsection{McNemar's Test on Discordant Pairs} \label{sec:mcnemar_comp} To compare classifier performance, we performed one-sided McNemar's tests \cite{McNemar_1947}. Let $n_{10}$ be the count of cases where MultiPhishGuard is correct and the alternative approach (the Chain-of-Thought method, the Single-Agent model, and the baseline RoBERTa-base) is incorrect, and let $n_{01}$ be the count of cases where the alternative approach is correct and MPG is incorrect. The tests are one-sided, with the hypothesis $n_{10} > n_{01}$. For $n_{10}+n_{01}>25$, we use McNemar's $\chi^2$ test using the \texttt{exact2x2} package \cite{fay2010two} with mid-p correction \cite{fagerland2013mcnemar}; otherwise we use McNemar's exact test using the \texttt{binom.test} \cite{R-base-binom.test} function. We adjusted $p$-values using Benjamini-Hochberg \cite{benjamini1995controlling}.\footnote{We control the false-discovery rate using Benjamini-Hochberg (BH) rather than a family-wise method (e.g. Bonferroni) because our primary aim is to screen for promising performance improvements across multiple benchmarks. BH offers greater power by limiting the expected proportion of false positives among significant results, rather than the stricter guarantee against any false positives.} The results of these tests are shown in Table \ref{tab:mcnemar_comp}.

\begin{table}
    \centering
\caption{One-sided McNemar's tests comparing MultiPhishGuard with CoT, SingleAgent, and baseline model across six datasets (\bm{$n_{10}$}: MultiPhishGuard correct, alternative incorrect; \bm{$n_{01}$} vice versa). We test \bm{$n_{10} > n_{01}$}.}
    \label{tab:mcnemar_comp}
    \begin{tabular}{p{1.35cm}lrrr} \toprule
\textbf{Dataset} & \textbf{MultiPhishGuard vs.} & \bm{$n_{10}$} & \bm{$n_{01}$} & \bm{$p$} \\ 
  \midrule
\multirow{3}{1.4cm}{Nigerian ($n=577$)} & CoT & 1 & 0 & 0.529 \\ 
   & SingleAgent & 0 & 0 & 1.000 \\ 
   & Baseline & 4 & 0 & 0.070 \\ 
   \midrule
\multirow{3}{1.4cm}{Nazario ($n=402)$} & CoT & 6 & 0 & 0.023 \\ 
   & SingleAgent & 4 & 0 & 0.070 \\ 
   & Baseline & 12 & 2 & 0.011 \\ 
   \midrule
\multirow{3}{1.4cm}{Enron ($n=1500)$}  & CoT & 263 & 4 & $<.001$ \\ 
   & SingleAgent & 305 & 3 & $<.001$ \\ 
   & Baseline & 62 & 37 & 0.011 \\ 
   \midrule
\multirow{3}{1.4cm}{SpamA. ($n=500)$}& CoT & 115 & 7 & $<.001$ \\ 
  & SingleAgent & 161 & 2 & $<.001$ \\ 
  & Baseline & 46 & 18 & $<.001$ \\ 
   \midrule
\multirow{3}{1.4cm}{CEAS-08 ($n=500)$}& CoT & 50 & 3 & $<.001$ \\ 
   & SingleAgent & 52 & 4 & $<.001$ \\ 
  & Baseline & 18 & 9 & 0.056 \\ 
   \midrule
\multirow{3}{1.4cm}{TREC-07 ($n=500)$}  & CoT & 28 & 0 & $<.001$ \\ 
   & SingleAgent & 23 & 0 & $<.001$ \\ 
   & Baseline & 18 & 8 & 0.036 \\ 
   \bottomrule
\end{tabular}
\end{table}

On the phishing email datasets (Nigerian and Nazario), MultiPhishGuard outperformed the alternatives on the majority of discordant comparisons, with the difference surpassing the threshold for significance in two cases: versus CoT and versus Baseline in the Nazario dataset. In the legitimate email datasets, MultiPhishGuard again dominated the discordant pairs and achieved statistically significant improvements over both Chain-of-Thought and Single-Agent across all corpora; it also significantly outperformed the RoBERTa-base model in the Enron, SpamAssassin, and TREC-07 datasets.

These results suggest that MultiPhishGuard consistently errs less often than the alternatives.

\subsubsection{Summary}

Our evaluation shows that MultiPhishGuard achieves high detection accuracy with a favorable balance between precision and recall, primarily driven by reduced false positive rates while maintaining high recall. As further detailed in the ablation studies (Section \ref{no_adversarial}), incorporating the adversarial agent appears to support improved robustness to evolving phishing strategies. Compared to CoT prompting, single-agent approaches, and RoBERTa-base, our approach achieves higher accuracy and improved explainability. Overall, these results suggest that MultiPhishGuard’s multi-agent, adaptive, and explainable design is effective across diverse phishing scenarios.

\subsection{Explanation Quality}

Beyond detection performance, we also evaluated the quality of the explanations generated by our system. As shown in Table \ref{tab:evaluation_reasons}, we measured the fluency of the explanations using perplexity, which was determined to be 25, indicating that the outputs are highly fluent. To assess the semantic consistency of the topics within the explanations, we calculated the topic coherence score, which stood at 0.35. Additionally, the Flesch Reading Ease Score was 41, suggesting that the explanations are moderately easy to read and comprehend for a general audience. Overall, these results highlight that MultiPhishGuard not only achieves strong discriminative performance but also provides clear, coherent, and accessible explanations, which we expect would lead to enhanced user trust and understanding.

\begin{table}
    \centering
\caption{Quality of model-generated explanations across approaches.}
\label{tab:evaluation_reasons}
    \begin{tabular}{lrrrr} \toprule
        \textbf{Approach} & \textbf{Perplexity} & \textbf{Topic Coherence} & \textbf{FRES} \\ \midrule
        \textit{MultiPhishGuard} & \textbf{25} & \textbf{0.35} & \textbf{41}  \\ 
        CoT & 47 & 0.28 & 21  \\
        No Explanation & 33 & 0.32 & 27  \\ 
        \bottomrule
    \end{tabular}
\end{table}

\subsection{Ablation Studies}

In our ablation studies, we evaluated the contribution of individual components in MultiPhishGuard by removing each component while keeping all others fixed, and comparing performance against the full model. We evaluate five ablated variants: (1) No URL Agent, (2) No Metadata Agent, (3) Static Weighting (no PPO), (4) No Adversarial Agent, and (5) No Explanation Simplifier. The results of the ablation studies are shown in Table \ref{tab:ablation_studies}. For completeness, per-dataset results are provided in Table \ref{tab:evaluation_ablation} and Section \ref{sec:mcnemar_abl}.

\begin{table*}
    \centering
    \rowcolors{3}{\rowgray}{white}
    \caption{Ablation study results showing the impact of removing individual components from MultiPhishGuard, aggregated across all datasets (all values in \%).}
    \label{tab:ablation_studies}
    \begin{tabular}{lrrrrrrr} \toprule
        \textbf{Variant} & 
        \textbf{Recall}
        & \textbf{Precision} 
        & \textbf{Accuracy}
        & \textbf{\foneb score}
        & \textbf{TNR}
        & \textbf{FPR}
        & \textbf{FNR} \\ \midrule
        \textit{MultiPhishGuard} & \textbf{99.80} & \textbf{92.26} & \textbf{97.89} & \textbf{95.88} & \textbf{97.27} & \textbf{2.73} & \textbf{0.20} \\
        No URL  & 97.96 & 73.83 & 90.95 & 84.20 & 88.67 & 11.33 & 2.04 \\ 
        No Metadata & 97.14 & 84.46 & 94.90 & 90.36 & 94.17 & 5.83 & 2.86  \\
        Static Weight & 99.18 & 84.80 & 95.43 & 91.43 & 94.20 & 5.80 & 0.82 \\
        No Adversarial & 98.77 & 84.31 & 95.17 & 90.97 & 94.00 & 6.00 & 1.23 \\
        \bottomrule
    \end{tabular}
\end{table*}

\subsubsection{No URL Agent}




As shown in Table \ref{tab:ablation_studies}, removing the URL agent led to a noticeable decline in detection performance. The absence of a dedicated URL analysis component significantly impaired the model’s ability to capture subtle phishing indicators embedded in hyperlinks---particularly those obfuscated through redirection or evasion techniques. This resulted in an increase in false positives and reduced overall detection robustness. These findings highlight the critical role of URL analysis in providing complementary insights that strengthen the system's ability to identify sophisticated phishing attempts. The experiment underscores the necessity of incorporating URL-based inspection in a multi-agent architecture to ensure comprehensive and resilient phishing detection.

\subsubsection{No Metadata Agent}

Removing the Metadata Agent led to a decline in detection performance, as the system became less capable of identifying subtle inconsistencies in email headers and sender details---signals that are often indicative of spoofing or other deceptive tactics. The absence of this component impaired the model’s ability to detect metadata-level phishing cues, such as forged sender addresses or manipulated header fields, which are frequently exploited in more sophisticated attacks. While the remaining agents continued to capture many phishing attempts through text and URL analysis, the overall robustness of the system was weakened. This experiment highlights the essential role of metadata inspection in enhancing detection accuracy and resilience, reinforcing the need for a multi-agent approach that integrates metadata analysis to address evolving phishing strategies.

\subsubsection{Static Weight}

We replaced the dynamic weight adjustment module with a static weighting scheme, where each agent's output was assigned a fixed, predetermined weight. From our previous experimental results, we observed that the URL agent had a greater impact on overall accuracy than the metadata agent. Removing the URL agent led to a more significant decline in detection performance, indicating its higher importance in phishing detection. Consequently, we assigned a weight of 0.3 to the text agent, 0.4 to the URL agent, and 0.3 to the metadata agent to better reflect their relative contributions to the final classification. This modification led to a decrease in detection performance, suggesting that the system struggled to balance the contributions of different agents effectively. Without the dynamic adjustment mechanism, the model lost the flexibility to tailor its weighting based on the unique characteristics of each email. Unlike the RL-based approach, which adapts in real-time to emphasize the most informative signals---be it from text, URL, or metadata---the static scheme imposed rigid constraints that limited the system’s responsiveness to varied phishing strategies. This rigidity makes the model more prone to misclassifications, particularly when phishing attempts relied heavily on one modality. The experiment underscores the importance of dynamic weight adaptation in enabling the system to make more informed and context-aware decisions, thereby enhancing its overall robustness and adaptability in phishing detection.

\subsubsection{No Adversarial Agent} \label{no_adversarial}

We removed the adversarial agent to assess its impact on the model's performance. This led to a decline in the system’s ability to detect more advanced and evasive phishing emails. The adversarial agent plays a key role in enhancing model resilience by generating deceptive, hard-to-detect phishing examples during training. Without this component, the model was exposed only to standard phishing patterns, limiting its capacity to generalize to more complex or novel attack strategies. Although the core detection capabilities remained intact, the model became more prone to misclassifying sophisticated phishing emails as legitimate. This highlights the critical role of adversarial training in fortifying the system against real-world threats and ensuring sustained effectiveness in the face of evolving phishing tactics.

\subsubsection{No Explanation Agent}
Finally, we removed the explanation simplifier agent to assess its impact on user comprehension and overall system usability. While this component does not directly influence detection accuracy, its absence significantly impaired the interpretability of the system’s outputs. The resulting explanations had a perplexity of 33, a topic coherence score of 0.32, and a Flesch Reading Ease score of 27---indicating moderate fluency but poor semantic organization and readability. Without the simplifier, the system continued to classify emails correctly but presented raw reasoning from individual agents in a fragmented and jargon-heavy manner. These explanations lacked coherence and were often difficult for non-expert users to interpret, reducing the system’s practical utility. Moreover, the absence of a unified, accessible explanation made it more time-consuming for analysts to piece together the rationale behind each decision. This experiment highlights the essential role of the explanation agent in translating technical outputs into clear, concise, and user-friendly narratives, which we expect would enhance user trust and support effective decision-making in phishing detection.

\subsubsection{McNemar's Test on Discordant Pairs} \label{sec:mcnemar_abl} We applied the same discordant pair analysis methodology described in Section \ref{sec:mcnemar_comp} to compare MultiPhishGuard against its ablated variants; the results are shown in Table \ref{tab:mcnemar_ablation}. 

\begin{table}
    \centering
\caption{One-sided McNemar's tests on discordant pairs for ablation variants (\bm{$n_{10}$}: MultiPhishGuard correct, ablated variant incorrect; \bm{$n_{01}$} vice versa) across six corpora. We test \bm{$n_{10} > n_{01}$}.}  
    \label{tab:mcnemar_ablation}
    \begin{tabular}{p{1.35cm}lrrr} \toprule
        \textbf{Dataset} & \textbf{MultiPhishGuard vs.} & \bm{$n_{10}$} & \bm{$n_{01}$} & \bm{$p$} \\ 
  \midrule
\multirow{4}{1.3cm}{Nigerian ($n=577$)} & NoURL & 0 & 0 & 1.000 \\ 
   & NoMetadata & 2 & 0 & 0.300 \\ 
   & StaticWeight & 0 & 0 & 1.000 \\ 
   & NoAdversarial & 1 & 0 & 0.571 \\ 
   \midrule
\multirow{4}{1.2cm}{Nazario ($n=402)$}   & NoURL & 19 & 1 & $<.001$ \\ 
   & NoMetadata & 24 & 0 & $<.001$ \\ 
   & StaticWeight & 7 & 1 & 0.050 \\ 
   & NoAdversarial & 9 & 0 & 0.004 \\ 
   \midrule
\multirow{4}{1.35cm}{Enron ($n=1500)$}  & NoURL & 45 & 23 & 0.007 \\ 
   & NoMetadata & 72 & 30 & $<.001$ \\ 
   & StaticWeight & 77 & 21 & $<.001$ \\ 
   & NoAdversarial & 69 & 25 & $<.001$ \\ 
   \midrule
\multirow{4}{1.2cm}{SpamA. ($n=500)$}  & NoURL & 119 & 10 & $<.001$ \\ 
   & NoMetadata & 66 & 19 & $<.001$ \\ 
  & StaticWeight & 37 & 13 & $<.001$ \\ 
 & NoAdversarial & 40 & 9 & $<.001$ \\ 
   \midrule
\multirow{4}{1.2cm}{CEAS-08 ($n=500)$} & NoURL & 112 & 6 & $<.001$ \\ 
   & NoMetadata & 11 & 6 & 0.210 \\ 
   & StaticWeight & 7 & 1 & 0.050 \\ 
 & NoAdversarial & 18 & 6 & 0.019 \\ 
   \midrule
\multirow{4}{1.2cm}{TREC-07 ($n=500)$} & NoURL & 21 & 0 & $<.001$ \\ 
   & NoMetadata & 1 & 2 & 0.955 \\ 
   & StaticWeight & 11 & 5 & 0.140 \\ 
   & NoAdversarial & 16 & 5 & 0.021 \\ 
   \bottomrule
    \end{tabular}
\end{table}

\begin{table*}
    \centering
    \rowcolors{5}{\rowgray}{white}
  \caption{Ablation study results: email classification metrics by dataset (all values in \%).}
    \label{tab:evaluation_ablation}
    \begin{tabular}{lrrrrrrrrrrrr}
    \toprule
    &\multicolumn{4}{c}{\textbf{Phishing}} & \multicolumn{8}{c}{\textbf{Legitimate}}\\\cmidrule(lr){2-5}\cmidrule(lr){6-13}
    & \multicolumn{2}{c}{\textbf{Nigerian}} & \multicolumn{2}{c}{\textbf{Nazario}} & \multicolumn{2}{c}{\textbf{Enron}} & \multicolumn{2}{c}{\textbf{SpamAssassin}} & \multicolumn{2}{c}{\textbf{CEAS-08}} & \multicolumn{2}{c}{\textbf{TREC-07}} \\\cmidrule(lr){2-3}\cmidrule(lr){4-5}\cmidrule(lr){6-7}\cmidrule(lr){8-9}\cmidrule(lr){10-11}\cmidrule(lr){12-13}
    \rowcolor{white}\textbf{Variant} 
    & \textbf{TPR} & \textbf{FNR} 
    & \textbf{TPR} & \textbf{FNR} 
    & \textbf{TNR} & \textbf{FPR} 
    & \textbf{TNR} & \textbf{FPR} 
    & \textbf{TNR} & \textbf{FPR} 
    & \textbf{TNR} & \textbf{FPR} \\ \midrule
    \textit{MultiPhishGuard} & \textbf{100.00} & \textbf{0.00} & \textbf{99.50} & \textbf{0.50} & \textbf{97.47} & \textbf{2.53} & \textbf{95.00} & \textbf{5.00} & \textbf{98.20} & \textbf{1.80} & 98.00 & 2.00 \\
    No URL & \textbf{100.00} & \textbf{0.00} & 95.02 & 4.98 & 96.00 & 4.00 & 73.20 & 26.80 & 77.00 & 23.00 & 93.80 & 6.20 \\
    No Metadata & 99.65 & 0.35 & 93.53 & 6.47 & 94.67 & 5.33 & 85.60 & 14.40 & 97.20 & 2.80 & \textbf{98.20} & \textbf{1.80}\\
    Static Weight & \textbf{100.00} & \textbf{0.00} & 98.01 & 1.99 & 93.73 & 6.27 & 90.20 & 9.80 & 97.00 & 3.00 & 96.80 & 3.20\\
    No Adversarial & 99.83 & 0.17 & 97.26 & 2.74 & 94.53 & 5.47 & 88.80 & 11.20 & 95.80 & 4.20 & 95.80 & 4.20\\
    \bottomrule
    \end{tabular}
\end{table*}

In the phishing email datasets (Nigerian and Nazario), MultiPhishGuard outperformed the ablated variants on the majority of discordant comparisons. The Nigerian set showed no significant differences, while on the Nazario set MultiPhishGuard significantly outperformed the NoURL, NoMetadata, and NoAdversarial variants. 

In the legitimate email datasets, MultiPhishGuard again dominated the discordant pairs and achieved statistically significant improvements over the ablated variants in all cases---except StaticWeight and NoMetadata on the CEAS-08 and TREC-07 sets. 

\subsubsection{Summary}

Overall, these findings indicate that the URL and Metadata Agents, the PPO-based weighting scheme, and the adversarial agent each contribute substantially to MultiPhishGuard's performance, and the Explanation Simplifier to its interpretability.

\subsection{Human Evaluation}

We conducted a human evaluation with a cybersecurity expert to assess the model’s explanations. The expert reviewed a subset of phishing emails and documented their reasoning, including key indicators and explanations. We compared these analyses with explanations generated by MultiPhishGuard, CoT, and an ablated variant using automated metrics (ROUGE \cite{chin2004rouge} and cosine similarity; see Appendix \ref{rationales_metrics}). This provides a preliminary assessment of alignment between model-generated explanations and expert reasoning in terms of clarity, coherence, and factual consistency.

As shown in Table \ref{tab:human_evaluations}, MultiPhishGuard achieved a ROUGE-1 score of 0.59 and a cosine similarity of 0.82, outperforming CoT (ROUGE-1: 0.45; cosine similarity: 0.70). Removing the explanation simplifier agent reduced performance (ROUGE-1: 0.42; cosine similarity: 0.64), suggesting that this component contributes to improved alignment with expert analyses. These results indicate that MultiPhishGuard produces explanations that are more coherent and semantically aligned with expert reasoning under these metrics.

\begin{table}
    \centering
    \caption{Automated evaluation metrics comparing expert analyses with different model explanations}
    \label{tab:human_evaluations}
    \begin{tabular}{lrr} \toprule
        \textbf{Approach} & \textbf{ROUGE-1} & \textbf{Cosine Similarity} \\ \midrule
         \textit{MultiPhishGuard} & \textbf{0.59} & \textbf{0.82}   \\
        CoT & 0.45 & 0.70  \\
        No Explanation & 0.42 & 0.64  \\
        \bottomrule
    \end{tabular}
\end{table} 

In addition to automated metrics, we conducted a qualitative analysis comparing the expert’s written rationale with MultiPhishGuard’s explanations. The model’s outputs often reflected similar phishing indicators identified by the expert—such as suspicious links and abnormal subject lines—and in some cases highlighted additional cues. For example, the system flagged the use of generic salutations such as ``Dear Customers’’ instead of addressing recipients by name.

Together with the automated scores (ROUGE-1: 0.59, cosine similarity: 0.82), these observations suggest that MultiPhishGuard can identify a range of phishing signals and provide explanations that are reasonably aligned with expert reasoning.

\section{Discussion}

Our results suggest that multi-agent LLM architectures with adaptive coordination represent a promising direction for phishing detection. Across multiple datasets, MultiPhishGuard achieves higher detection accuracy, lower false positive rates, and improved robustness compared to strong research baselines.

Ablation results suggest these improvements are driven by its multi-agent architecture, where specialized agents---focused on email text, URL structure, and metadata---analyze different facets of an email before synthesizing insights through a dynamically optimized ensemble mechanism. Another key feature is the use of reinforcement learning via PPO to dynamically adjust agent weights based on email-specific characteristics. Unlike prior work relying on static or heuristic weighting, this dynamic approach allows the system to emphasize the most informative modalities, which may be beneficial in real-world deployment scenarios where phishing cues and attack strategies change over time.

Another notable contribution is the system’s explainability. By incorporating an explanation simplifier agent, MultiPhishGuard generates clear, non-technical rationales for its decisions. Human evaluation and automated metrics indicate that these explanations are broadly consistent with expert reasoning and are more readable and coherent than those produced by CoT or ablated variants. This suggests that the framework may be particularly relevant for user-facing and high-stakes applications where transparency and auditability are required, including settings subject to regulatory constraints (e.g., GDPR).

To enhance robustness, MultiPhishGuard incorporates an LLM-based adversarial training module that generates subtle phishing and legitimate email variants. By exposing the model to these challenging cases during training, the approach is intended to improve robustness to diverse and evolving attack patterns. Ablation results suggest that removing this component reduces performance on more subtle or ambiguous phishing examples, indicating its contribution to handling difficult cases.

The system’s modular design may support practical deployment flexibility. For example, organizations could, in principle, enable only the text agent to reduce reliance on potentially sensitive metadata. This flexibility may be useful in privacy-sensitive environments, although it involves trade-offs in detection performance, as suggested by our ablation results.

Looking forward, the architecture could be extended to include additional specialized agents (e.g., one focused on attachment analysis). MultiPhishGuard illustrates a design pattern for scalable, interpretable multi-agent systems that may generalize to broader cybersecurity domains such as malware detection and malicious website classification. For practitioners, it suggests how modular, explainable, and adaptive LLM-based frameworks could be applied in real-world settings where effectiveness and accountability are important.

In summary, MultiPhishGuard integrates multi-modal analysis, dynamic optimization, and interpretability within a unified phishing detection framework. It demonstrates how these elements can be combined in a coherent system and provides a foundation for future work on adaptive and explainable LLM-based cybersecurity tools.

\subsection{Limitations}

Our work has several limitations that warrant further investigation. First, the interpretability of the generated rationales is difficult to evaluate due to the lack of ground-truth explanations in existing datasets. As a result, we rely on indirect metrics such as readability, coherence, perplexity, and limited human assessment. Second, our evaluation is constrained by the availability of open-source phishing detection models and datasets. We were able to compare against only one recent state-of-the-art model with publicly available code, and broader comparisons across additional high-quality benchmarks would strengthen the evaluation. Third, we do not systematically assess how the framework generalizes across different model families or cost–performance trade-offs. Finally, although our system includes an explanation simplifier, we do not conduct user studies to evaluate how explanations are interpreted by users with different expertise. Understanding how these explanations support decision-making remains an important direction for future work.

\subsection{Ethical Considerations}

We address several ethical and safety considerations in the design and evaluation of our system. All email data is anonymized and handled in accordance with privacy regulations to prevent disclosure of personally identifiable information. We also ensure that generated explanations do not expose sensitive decision-making processes or provide actionable guidance for evading detection.

To evaluate robustness, we use an adversarial agent to generate both phishing and legitimate email variants using GPT-4o, reflecting the potential misuse of LLMs for malicious content. These emails are created in a secure, controlled environment, used solely for internal testing, and are not released to prevent misuse.

We further address the risk of LLM hallucination by constraining agents to operate within their respective modalities and requiring the explanation simplifier to ground outputs in agent-provided reasoning. Combined with prompt design and output constraints, this reduces unsupported claims and helps ensure accurate, trustworthy outputs.

\section{Conclusion}
In this paper, we introduced MultiPhishGuard, a phishing detection framework that leverages an LLM-based multi-agent system to support both detection performance and interpretability. By integrating specialized agents for text, URL, and metadata analysis, and dynamically adjusting their contributions using PPO, our approach achieves higher performance than strong research baselines, including single-agent methods and Chain-of-Thought prompting. The adversarial training module is designed to improve robustness to subtle and ambiguous phishing cases during training. Experimental results indicate that MultiPhishGuard achieves high detection accuracy across the evaluated datasets (97.89\%), while maintaining low false positive and false negative rates. Additionally, the inclusion of an explanation simplifier agent provides clear, user-friendly rationales for its decisions, addressing a key limitation in model transparency.

Overall, MultiPhishGuard provides an adaptive and interpretable approach to phishing detection. These results suggest that combining multi-agent architectures with dynamic weighting and adversarial training is a useful direction for developing more robust and transparent cybersecurity systems.

\bibliographystyle{IEEEtran}
\bibliography{References}

\clearpage
\appendices

\section{Full Agent Prompts} \label{prompts}

\begin{figure}[h]
    \centering
   \begin{tikzpicture}[
  box/.style={
    minimum width=1cm,
    inner sep=6pt,
    align=left, 
    rounded corners=8pt,
    fill=black!5
  }
  ]
    \node[box] (A) {%
\begin{minipage}{\dimexpr\columnwidth-12pt}

\setlength{\parskip}{1.5ex}

\ttfamily\small
You are a cybersecurity expert specializing in phishing, with a particular focus on URLs within emails. Your task is to carefully examine every URL in the email and determine whether it exhibits suspicious characteristics or signs of obfuscation, such as resembling forged bank sites, using unknown domains, or appearing unusually shortened. Do not analyze the email text or metadata, only focus on the URLs. Provide your judgment on whether the email is `Phishing' or `Legitimate', along with a confidence score between 0 and 1 and a clear, concise explanation of your reasoning. Output your result in JSON format as: {`verdict': `Phishing' or `Legitimate', `confidence': 0-1, `reasons': `...'}

\end{minipage}
};
\end{tikzpicture}%
    \caption{URL Agent Prompt}
    \label{fig: URL agent prompt}
\end{figure}

\begin{figure}[h]
\centering
\begin{tikzpicture}[
  box/.style={
    minimum width=1cm,
    inner sep=6pt,
    align=left, 
    rounded corners=8pt,
    fill=black!5
  }
  ]
    \node[box] (A) {%
\begin{minipage}{\dimexpr\columnwidth-12pt}

\setlength{\parskip}{1.5ex}

\ttfamily\small
You are a cybersecurity expert specializing in phishing, with a particular focus on email metadata. Your task is to scrutinize the provided email header—including the subject, sender address, reply-to, return-path, and received fields—for any signs of forgery, anomalies, or suspicious indicators. Do not analyze email text or URLs, only focus on the metadata. Provide your judgment on whether the email is 'Phishing' or 'Legitimate', along with a confidence score between 0 and 1 and a clear, concise explanation of your reasoning. Output your result in JSON format as: {`verdict': `Phishing' or `Legitimate', `confidence': 0-1, `reasons': `...'}
\end{minipage}
};
\end{tikzpicture}%
\caption{Metadata Agent Prompt}
    \label{fig: metadata agent prompt}
\end{figure}

\begin{figure}
\centering
\begin{tikzpicture}[
  box/.style={
    minimum width=1cm,
    inner sep=6pt,
    align=left, 
    rounded corners=8pt,
    fill=black!5
  }
  ]
    \node[box] (A) {%
\begin{minipage}{\dimexpr\columnwidth-12pt}
\ttfamily
\setlength{\parskip}{1.5ex}
\footnotesize

You are an expert adversarial email generator. Your objective is to produce a variant of the provided email that maintains the original meaning and structure while incorporating subtle modifications designed to bypass the phishing detectors. Depending on the type of email provided, follow the corresponding instructions:
        
        For phishing emails:\vspace{-1.5ex}
        \begin{enumerate}
        \item Synonym Substitution: Replace keywords with synonyms (e.g., ``verify'' → ``confirm'', ``account'' → ``profile'', ``free'' → ``no money is needed'') so that the literal expression changes while the meaning remains intact.
        \item Sentence Rewriting: Alter the sentence structure without changing the underlying message (e.g., transform ``Update your account immediately'' into ``Please refresh your account details at your earliest convenience''). Add decoy sentences about customer support/legitimate services. Remove overt threat indicators while maintaining urgency.
        \item Content Modification: Add or remove words and phrases as needed; for example, insert a neutral sentence like ``We hope this email serves you well'' or omit less critical content, to change the text's composition.
        \item Homoglyph Replacement: Substitute characters with similar-looking counterparts (e.g., replace the letter ``a'' in ``paypal.com'' with a Cyrillic ``a'' to disguise URLs while retaining their recognizable form.
        \item Polymorphic Variation: Modifying aspects such as the subject line, sender information, or overall format, thereby simulating a diverse range of phishing attack styles.
        \end{enumerate}
      
        For legitimate emails:
        \vspace{-1.5ex}

        \begin{enumerate}
        \item Subtle Suspicious Modifications: Modify the email in ways that make it appear more ambiguous or borderline suspicious (e.g., incorporate slightly urgent language or modify the subject line) without compromising its inherently benign intent.  
        \item Synonym Substitution and Sentence Rewriting: Use similar techniques as above but ensure that the overall message remains authentic and professional, even if the modifications introduce elements that could potentially confuse detection systems.  
        \item Content Enhancement: Optionally insert additional phrases that mimic some characteristics of phishing emails (e.g., ambiguous urgency or formatting cues), while still maintaining the legitimacy of the email.  
        \item Polymorphic Variation: Adjust non-critical elements like the layout or minor stylistic details to introduce natural variability without altering the email's genuine nature.
        \end{enumerate}
        
        Output Requirements:\vspace{-1.5ex}

        \begin{itemize}
        \item For phishing emails, the final variant should retain the malicious intent and target brand while evading detection.  
        \item For legitimate emails, the final variant should remain clearly benign and professional, yet include subtle modifications that challenge the detector.  
        \item Provide only the final modified email text and do not disclose the modification details.
        \end{itemize}
        \end{minipage}
        };
    \end{tikzpicture}%
    \caption{Adversarial Agent Prompt}
    \label{fig: adversarial agent prompt}
\end{figure}

\FloatBarrier 
\section{Example Adversarial Email Variant} \label{adversarial_email}

\begin{figure}[H]
\centering
\begin{tikzpicture}[
  box/.style={
    draw, 
    minimum width=1cm,
    inner sep=6pt,
    align=left, 
    rounded corners=6pt
  },
  node distance=.8cm
  ]

  \node[box] (A) {%
\begin{minipage}{\dimexpr\columnwidth-12pt\relax}
    
\sffamily
\setlength{\parskip}{2ex}

\textbf{From:} {``Monkey Support Notification'' info@creditloiuse.com}\\
\textbf{Subject:} Important Password Validation\\
\textbf{To:} jose@monkey.org\\
\textbf{Date:} Wed, 10 Jan 2024 16:00:51 -0800

      Monkey Support Notification

      ACTION REQUIRED: Monkey (jose@monkey.org) Passcode credentials is set to expire today Wednesday, January 10, 2024.

      Same registered password must be used below.

      RETAIN CREDENTIALS \\\fakelink{\underline{https:bafkreigdpekci4il2gjagofvg7irzyjqqkam2hz36lrx3ms7}\\\underline{fwfpvxdjjq.ipfs.dweb.link/\#jose@monkey.org}}

      Note: It is important you update your email passcode, failure to do this will get your account locked out and email will not be accessible.

      Update your email preferences to choose the types of emails you receive, or you can unsubscribe from all future emails.
    \end{minipage}
  };
\end{tikzpicture}
        
 
\caption{Original phishing email.}
\label{fig: original_email}
\end{figure}

\begin{figure}[H]
\centering
\begin{tikzpicture}[
  box/.style={
  draw, 
    minimum width=1cm,
    inner sep=6pt,
    align=left, 
  },
  node distance=.8cm,
  rounded corners=8pt
  ]

\sffamily 
\setlength{\parskip}{2ex}

  \node[box] (A) {%
  \begin{minipage}{\dimexpr\columnwidth-12pt\relax}

      \setlength{\parskip}{2ex}

      \textbf{From:} ``Monkey Assistance Notice'' info@creditloiuse.com\\
        \textbf{Subject:} Vital Password Confirmation\\
        \textbf{To:} jose@monkey.org\\
        \textbf{Date:} Wed, 10 Jan 2024 16:00:51 -0800

      MONKEY NOTICE: The passkey linked to your profile (jose@monkey.org) will reach its expiration this Wednesday, January 10, 2024.

      Please ensure you are utilizing the same registered password using the link below.

      ACCESS CREDENTIALS: \\ 
      \fakelink{\underline{https://bafkreigdpekci4il2gjagofvg7irzyjqqkam2hz36lrx3m}\\\underline{s7fwfpvxdjjq.ipfs.dweb.link/\#jose@monkey.org}}

      Please be advised: It's crucial to confirm your email password soon to avoid any disruptions in accessing your email.

      You are welcome to update your email settings for receiving different categories of emails, or you can opt out from all subsequent communications.
    \end{minipage}
  };
\end{tikzpicture}
\caption{Adversarial agent--generated variant.}
\label{fig: adversarial_email}
\end{figure}


\FloatBarrier
\section{Evaluation Metrics}
\subsection{Evaluation Metrics for Classification} \label{classification_metrics}
\begin{itemize}
    \item \textbf{True Positives (TP):} The number of phishing emails correctly identified as phishing.
    \item \textbf{True Negatives (TN):} The number of legitimate emails correctly classified as legitimate.
    \item \textbf{False Positives (FP):} The number of legitimate emails mistakenly classified as phishing.
    \item \textbf{False Negatives (FN):} The number of phishing emails incorrectly classified as legitimate.
\end{itemize}
Using these fundamental values, we calculate the following evaluation metrics:
\begin{itemize}
    \item \textbf{Recall:} Measures how well the model identifies phishing emails.
    \[\text{Recall} = {TPR} = \frac{TP}{TP + FN}\]
    A high recall ensures that most phishing emails are detected, reducing the chances of missed attacks.
    \item \textbf{Precision:} Measures how many of the emails classified as phishing are actually phishing.
    \[\text{Precision} = \frac{TP}{TP + FP}\]
    A higher precision means fewer legitimate emails are mistakenly flagged as phishing, reducing false alarms.
    \item \textbf{Accuracy:} Measures the overall correctness of the model's classifications. 
    \[\text{Accuracy} = \frac{TP + TN}{TP + TN + FP + FN}\]
    A high accuracy indicates strong overall performance, but it may not be reliable when the dataset is imbalanced.
    \item \textbf{\fone Score:} The harmonic mean of precision and recall, balancing both metrics to provide a single performance measure.
    \[\text{\fone score} = 2 \times \frac{\text{Precision} \times \text{Recall}}{\text{Precision} + \text{Recall}}\]
    A high \fone score indicates that the model performs well in both phishing detection and avoiding false alarms.
    \item \textbf{True Negative Rate (TNR):} Measures the proportion of legitimate emails correctly classified.
    \[\text{TNR} = \frac{TN}{TN + FP}\]
    A high TNR ensures that the system does not mistakenly flag too many legitimate emails as phishing.
    \item \textbf{False Positive Rate (FPR):} Measures the proportion of legitimate emails incorrectly classified as phishing.
    \[\text{FPR} = \frac{FP}{FP + TN}\]
    A lower FPR is desirable, as it reduces unnecessary phishing alerts and minimizes disruptions to users.
    \item \textbf{False Negative Rate (FNR):} Measures the proportion of phishing emails incorrectly classified as legitimate.
    \[\text{FNR} = \frac{FN}{TP + FN}\]
    A low FNR is critical, as it minimizes the risk of phishing emails bypassing detection and reaching users.
\end{itemize}

\subsection{Evaluation Metrics for Rationales} \label{rationales_metrics}
\begin{itemize}
    \item \textbf{Perplexity (PPL):} Perplexity \cite{jelinek1977perplexity} is a widely used metric in natural language processing to evaluate the fluency and predictability of text generation. It quantifies how well a language model predicts the next word in a sequence, with lower perplexity indicating more coherent and fluent explanations.
    \[\text{PPL} = \exp\left( -\frac{1}{N} \sum_{i=1}^N \log p(w_i) \right)\]
    \( N \) is the number of words in the explanation, and \( p(w_i) \) is the probability assigned to each word. A lower perplexity score suggests that the explanation is more readable and natural, while a higher perplexity indicates disjointed or unnatural phrasing.
    \item \textbf{Topic Coherence:} Topic Coherence \cite{rosner2014evaluating} evaluates the semantic consistency of topics within the explanations, ensuring they are interpretable and logically structured. It measures the degree of semantic similarity among words within a topic, reflecting the interpretability of the explanations. Higher coherence scores indicate that the explanations are more logically structured and easier to understand. This ensures that the explanations are not only accurate but also presented in a manner that is coherent and user-friendly.
    \item \textbf{Flesch Reading Ease Score (FRES):} The Flesch Reading Ease Score \cite{flesch1979write} is a widely used metric to evaluate the readability of a text. Developed by Rudolf Flesch, it assigns a score between 0 and 100, with higher scores indicating easier readability. The formula for calculating this score is:
    \begin{multline*}
    FRES = 206.835 - 1.015 \times (\frac{\text{Total Words}}{\text{Total Sentences}})\\ 
    - 84.6 \times (\frac{\text{Total Syllables}}{\text{Total Words}})
    \end{multline*}
    In the context of evaluating explanations, applying the Flesch Reading Ease Score can help assess how easily readers can comprehend the provided reasons. A higher score suggests that the explanation is straightforward and accessible, while a lower score may indicate complexity that could hinder understanding. For instance, explanations laden with technical jargon, lengthy sentences, or complex words are likely to yield lower readability scores, signaling the need for simplification to enhance clarity.
    \item \textbf{ROUGE (Recall-Oriented Understudy for Gisting Evaluation):} ROUGE \cite{chin2004rouge} quantifies the overlap of n-grams between the candidate text (MultiPhishGuard’s explanation) and the reference text (the expert’s analysis), allowing us to measure how much of the expert’s key content is captured by our model. We focus on ROUGE-1, which computes unigram recall:
    \[
    \text{ROUGE-1 Recall} = \frac{\sum_{w \in R} \min(\text{Count}(w, C), \text{Count}(w, R))}{\sum_{w \in R} \text{Count}(w, R)}
    \]

    where \(\text{Count}(w, C)\) and \(\text{Count}(w, R)\) are the frequencies of word \(w\) in the candidate and reference texts, respectively. A high ROUGE-1 score indicates that the system’s output includes a large proportion of the expert’s important unigrams, which is desirable because it shows that the generated explanation covers the essential content.
    We choose ROUGE-1 over higher-order variants (ROUGE-2, ROUGE-L, etc.) because expert explanations and LLM-generated outputs often differ in phrasing and sentence structure. Unigram overlap provides a robust, style-agnostic measure of content similarity, whereas bigram or sequence-based metrics can unfairly penalize legitimate rewordings and stylistic variations. By using ROUGE-1, we ensure a balanced evaluation that emphasizes the presence of critical terms without over-penalizing differences in natural human versus model-generated language.
    \item \textbf{Cosine Similarity:} Cosine similarity measures the semantic similarity between two texts by converting them into vector representations (e.g., using sentence embeddings) and computing the cosine of the angle between these vectors. It is defined as:
    \[
    \text{Cosine Similarity} = \frac{\mathbf{A} \cdot \mathbf{B}}{\|\mathbf{A}\|\|\mathbf{B}\|}
    \]
    where \(\mathbf{A}\) and \(\mathbf{B}\) represent the expert’s and the model’s explanation embeddings, respectively. Values close to 1 indicate that the two texts are semantically very similar, meaning the system’s explanation has captured the overall meaning and context of the expert’s analysis. Unlike ROUGE---which focuses on exact word overlap---cosine similarity captures deeper, conceptual alignment. This allows our explanations to retain semantic similarity and contextual relevance to the expert’s intent, even when the wording differs.
\end{itemize}

\end{document}